\documentclass[review]{elsarticle}

\usepackage{enumitem}
\setlist[enumerate,1]{label={\normalfont(\roman*)}}
\makeatletter
\newcommand*{\rmn}[1]{\expandafter\@slowromancap\romannumeral #1@}
\makeatother
\usepackage{lineno,hyperref}
\usepackage{graphicx}	% Including figure files
\usepackage{amsmath}	% Advanced maths commands
\usepackage{amssymb}	% Extra maths symbols
\usepackage{multirow}
\usepackage{tabularx}
\usepackage[para,online,flushleft]{threeparttable}
\usepackage[T1]{fontenc}
\usepackage{ae,aecompl}
\usepackage{gensymb}
\usepackage{xfrac}
\usepackage{siunitx}
\usepackage{natbib}
\usepackage{xcolor}
\modulolinenumbers[5]

\journal{New Astronomy}

\begin{document}

\begin{frontmatter}

\title{NGC 2366 : An optical search for possible supernova remnants \tnoteref{A search for SNR in NGC 2366}}
\tnotetext[mytitlenote]{Fully documented templates are available in the elsarticle package on \href{http://www.ctan.org/tex-archive/macros/latex/contrib/elsarticle}{CTAN}.}

%% Group authors per affiliation:
\author{E. N. Ercan\fnref{Ercan}}
\address{Department of Physics, Bogazici University, Istanbul, 34342, Turkey}
\fntext[myfootnote]{ercan@boun.edu.tr}

\author{E. Aktekin\fnref{Aktekin}}
\address{Currently Self-Employed, 1728 Street, Muratpasa, Antalya, 07058, Turkey}
\fntext[myfootnote]{aktekinebru@gmail.com}

\begin{abstract}
 The results of an optical search for supernova remnants (SNRs) in the nearby irregular galaxy NGC 2366 are presented. We took interference filter images and collected spectral data in three epochs with the f/7.7 1.5 m Russian Turkish Telescope (RTT150) at T\"{U}B\.{I}TAK National Observatory (TUG) located in Antalya, Turkey. The continuum-subtracted H$\alpha$ and continuum-subtracted [S\,{\sc ii}]$\lambda \lambda$6716, 6731 images and their ratios were used for the identification of SNRs. With [S\,{\sc ii}]/H$\alpha$ $\geq$ 0.4 criteria, four possible SNR candidates were identified in NGC 2366 with [S\,{\sc ii}]/H$\alpha$ ratios of $\sim$(0.68, 0.57, 0.55 and 0.75), H$\alpha$ intensities of $\sim$(2.10, 0.36, 0.14, 0.11)$\times 10^{-15}$ erg cm$^{-2}$ s$^{-1}$ [S\,{\sc ii}]$\lambda$6716/$\lambda$6731 average flux ratios of $\sim$(1.01  and 1.04), electron densities of $N_{\rm e}$ $\sim$(582  and  513) cm$^{-3}$ and [O\,{\sc iii}] $\lambda$5007/H$\beta$ $\lambda$4861 $\sim$(3.6  and 2.6) line ratio values are obtained for two of the SNR candidates. A shock velocity $V_{\rm s}$ of 80 $\leq$ $V_{\rm s}$ $\leq$ 100 km s$^{-1}$ is reported. The spectral parameters are obtained for the first time for these possible SNR candidates. The locations of the four SNRs obtained here are found to be consistent with  optical and radio  results reported so far. One of the sources categorised earlier by \textit{XMM-Newton} observations as an extended X-ray source position is found to be consistent with one of four possible SNR candidates reported here.
\end{abstract}

\begin{keyword}
\texttt{ISM: supernova remnants -- galaxies: irregular -- galaxies: individual: NGC 2366 -- galaxies.}

\end{keyword}

\end{frontmatter}

\section{Introduction}

Supernova explosions and their final dispersion of ejected material have the effect of supplementing the interstellar medium (ISM) with the material produced in stellar interiors and therefore  supernova remnants (SNRs) are important for many of the theories of the ISM. Due to the strong photoionisation flux of the central hot star or stars, sulfur is in the form of S$^{++}$ in H\,{\sc ii} regions, where the [S\,{\sc ii}]/H$\alpha$ ratio is typically $\approx$ 0.1 -- 0.3. Outside an H\,{\sc ii} region, there are not sufficient energetic free electrons to excite S$^{+}$ and to give rise to the forbidden-line ([S\,{\sc ii}]$\lambda$6716, 6731) emission. Almost all discrete emission nebulae outside H\,{\sc ii} regions showing [S\,{\sc ii}]-emission are shock-heated and they are probably SNRs and are characterized by [S\,{\sc ii}]/H$\alpha$ $\geq$ 0.4.

Several early studies discussed the reasons and importance of searching SNRs in nearby galaxies \citep[see e.g.][]{matonickandfesen1997, pannutietal2000, pannutietal2002} and in our Galaxy \citep[see e.g.][]{mavromatakisetal2002, fesenetal1997, fesenetal2008}. Although there are a large number of Galactic SNRs, the interstellar extinction and uncertain distances cause selection effects. However, in extragalactic samples, these difficulties are much less. All the SNRs in a galaxy are at the same distance from us. Assuming that we know the distance of this galaxy, the properties of these SNRs can be compared directly. In addition, in an extragalactic survey the foreground extinction is generally low such that the relative positions of SNR samples can be determined precisely. The distributions of the SNRs relative to the H\,{\sc ii} regions in the spiral arms of the Galaxy are calculated using the location of SNRs. Possible SNR progenitors have been searched from these distributions as indicated by \citet{matonickandfesen1997} and  \citet{blairandlong1997, blairandlong2004}. Extragalactic searches for SNRs were first carried out for the Magellanic Clouds by \citet{mathewsonandclarke1973}. They were the first to use the [S\,{\sc ii}]/H$\alpha$ emission line ratios for optical identification of SNRs. \citet{blairetal1981, smithetal1993, blairandlong1997, blairandlong2004}, \citet{sonbasetal2009, sonbasetal2010} and \citet{leonidakietal2013} have already discussed the motivation for observing SNRs in nearby galaxies based on the same emmision-line ratio criterion. Radio searches for extragalactic SNRs have been conducted by \citet{laceyetal1997, laceyandduric2001, hymanetal2001}. SNR surveys have also been carried out at optical, radio, and X-ray wavelengths by \citet{pannutietal2000, pannutietal2002, pannutietal2007}.

In this work, we searched for SNRs in the nearby dwarf,  irrregular  galaxy NGC 2366 using the criterion [S\,{\sc ii}]/H$\alpha$ ratio $\geq$ 0.4.  It is described as a ``cometary'' galaxy by \citet{looseandthuan1986}. \citet{karachentsevaetal1985} reported its distance to be $D$=3.44 Mpc, which is consistent with \citet{tikhonovetal1991} who reported $D$=3.4 Mpc. The optical and radio studies by \citet{Kennicuttetal1980} showed evidence of very large H\,{\sc ii} regions in NGC 2366. From the emission-line spectrum of this galaxy, it was found to be exceptional due to its excessively high degree of excitation together with a very low reddening. The authors attributed these features to a very low metal abundance, suggesting the existence of an exceptionally energetic source of ionization. Its 21 cm distinctive spectrum and large scale H\,{\sc ii} regions are reminiscent of the more distant narrow-lined Markarian galaxies and isolated extragalactic H\,{\sc ii} regions. \citet{yangetal1994} have detected 5 H$\alpha$ sources with broad velocity profiles compared to those of their surrounding H\,{\sc i} regions. This galaxy is also classified as a dwarf blue galaxy, showing indications of spiral arm structure at several wavelengths \citep[]{devaucouleursetal1991}. It is dominated in H$\alpha$ by the huge extragalactic H\,{\sc ii} region in its south west part  as indicated by \citet{chuandkennicutt1994}. \citet{thuanandizotov2005} provided its $V$ and $I$ photometric results by resolving stars with Wide Field Planetary Camera 2 images obtained from the Hubble Space Telescope (HST). Their 23.65 magnitude limit of the red giant branch stars with ages of $>$ 1 Gyr provided another distance estimate of 3.42 $\pm$ 0.15 Mpc which shows a good agreement with the aforementioned values. Early abundance measurements have been reported for this galaxy by several studies \citep{masegosaetal1991, gonzalez-delgadoetal1994, noeskeetal2000}. \citet{royetal1991, royetal1992} suggested the existence of an expanding supernova bubble in NGC 2363 which is, in fact, a giant H\,{\sc ii} region within NGC 2366. Its early radio observations were reported by \citet{kleinandgrave1986} at $\lambda$6.3 and $\lambda$2.8 cm. The authors measured the  whole galaxy (in low resolution) and  obtained 10$\pm$1 mJy flux at 6.3 cm. Their measurement of only the giant H\,{\sc ii} region at 2 $''$ resolution yielded 8.5 $\pm$ 0.28 mJy flux at $\lambda$6.3 cm. The difference reported in their  spectral indices between the entire galaxy and the giant H\,{\sc ii} region was attributed to the probable dominant thermal emission across the galaxy. Also \citet{chomiukandwilcots2009} reported radio emission of NGC 2366 from VLA continuum data. They diagnosed a number of discrete sources and categorized them as possible SNRs, H\,{\sc ii} regions, and/or background radio galaxies. Their findings were in good agreement with various studies and with four SNR candidates in NGC 2366 we report here (see  our Fig.~\ref{fig:fig1}).

Use of the multiwavelength observations is a great help when trying to understand  the characteristics and  physical parameters of SNRs as has been pointed out by various studies \citep[see e.g.][]{matonickandfesen1997, leeandlee2014, thuanetal2014}.

In the earliest photometric studies, NGC 2366 is described as a nearby galaxy with intense and rare star forming bursts characteristic for evolution of late-type dwarf galaxies \citep{searleetal1973}. \citet{jaskotandoey2013} and  \citet{michevaetal2017} also reported this galaxy in their study of the most extreme so-called Green Peas galaxies with the highest [O\,{\sc iii}]/[O\,{\sc ii}] ratios. Their [O\,{\sc iii}]$\lambda\lambda$5007, 4959/$\lambda$4363 ratios yielded electron temperatures of $T_{\rm e}$ $\sim$ 15 000 K while their [S\,{\sc ii}]$\lambda$6716/$\lambda$6731 ratios  provided  electron densities of $N_{\rm e}$ = 100--1000 cm$^{-3}$ in NGC 2366. $T_{\rm e}$ $\sim$ 15 000 K was reported for nebular emission by several authors \citep{gonzalez-delgadoetal1994, sokaletal2016}. \citet{gonzalez-delgadoetal1994} also reported an electron density  of $N_{\rm e}$ = 235 $\pm$ 41 cm$^{-3}$.  Recently, \citet{vuceticetal2019} reported their photometric observational results for NGC 2366  and suggested the presence of 67 possible H\,{\sc ii} regions  and two  optical SNR candidates.

\citet{thuanetal2014} report the \textit{XMM-Newton} results of NGC 2366 indicating the  existence of  two faint X-ray point sources and two faint extended sources. These authors assigned one of these point sources to a possible background active galactic nucleus (AGN), while the other source was found to be coincident with a very luminous star and a compact H\,{\sc ii} region. Their two faint extended sources were reported to be possibly  associated with massive H\,{\sc ii} regions. Another study by \citet{stevensandstrickland1998} was the first \textit{ROSAT} PSPC survey of Wolf-Rayet (WR) galaxies, including NGC 2366. In their study, NGC 2366 was described as a dwarf starburst barred spiral galaxy with an AGN. NGC 2366's \textit{ROSAT} PSPC X-ray position was found to be RA(J2000)=$07^{h} 28^{m}24^{s}$,  Dec.(J2000)=$\ang[angle-symbol-over-decimal]{69;11;00}$. These authors traced out WR galaxies and made a possible detection of WR stars in NGC 2363 as was earlier reported by \citet{drissenetal1993}. \citet{gonzalez-delgadoetal1994} presented an optical study of NGC 2363 and found broad emission lines at $\lambda$4660 and $\lambda$5810 {\AA} indicating the presence of WR stars. They reported a combined mass of these stars of 3.4 $\times$ 10$^{5}$ \(M_\odot\), suggesting the existence of a strong outflow, possibly due to the blow-out of a superbubble. Radio observations  with lower resolution suggested the emission from NGC 2366 as a whole should be dominated by thermal emission  \citep{yangetal1994}. \citet{stevensandstrickland1998}'s conclusion was that they detected an extended region, however rather with ambiguity. They provided an optical image superimposed with X-ray contours, pointing to a faint extended  X-ray emission coming from the nearby star-forming region in NGC 2363 with a contour level of 2.8 $\times$ 10$^{-3}$ count s$^{-1}$ arcmin$^{-2}$. They also found an X-ray point source that might be  associated with NGC 2366. The X-ray point  source was too faint to obtain a proper spectral fit, but in order to provide its X-ray luminosity they fitted a spectral model with a temperature of $kT$ = 0.52 keV, a metallicity of 0.1 $Z_\odot$, and a column density of 4.52 $\times$ 10$^{20}$ cm$^{-2}$. With this fit, they obtained an unabsorbed X-ray luminosity of 6.61 $\times$ 10$^{37}$ erg s$^{-1}$. However, they argued that it could be an underestimation of the real X-ray luminosity in case the actual  galactic column density was higher. Therefore, they did not report best-fit values for their analysis. We will  compare our possible SNR candidates locations obtained in this work with their corresponding X-ray positions.

Here, we report a detailed study of optical CCD imaging and spectroscopic analysis of NGC 2366. Since there  were no X-ray data available other than those summarised  here  obtained by several authors as indicated above, we could not perform any X-ray data analysis and therefore could not report them here but rather  we made some comparison with possible SNR locations obtained in this work and the early limited  X-ray results published so far.

Our work is structured as follows. The description of the observations and data reduction is given in Section \ref{Obs}. Our discussion and conclusions can be found in Section \ref{Conclusions}. The fundamental physical properties of NGC 2366 are given in Table~\ref{tab:Table1}.

\begin{table*}
\centering
 \caption{NGC 2366 properties.}
 \label{tab:Table1}
 \begin{tabular}{@{}ccc@{}}
\hline
     Parameters & Value & References \\
     \hline
Morphological Type & IB(s)m  & \citealt{Herrmannetal2016}\\
\multirow{2}{*}{Position (J2000)}& $RA(J2000)$ : $07^{h} 23^{m} 51^{s}$.85  & \multirow{2}{*}{\citealt{Cottonetal1999}}\\
&$Dec.(J2000)$: Dec.(J2000)=$\ang[angle-symbol-over-decimal]{69;12;31.10}$ & \\
Redshift&0.000330& \citealt{Patureletal2002}\\
Distance& 3.44 Mpc & \citealt{tolstoyetal1995} \\
Diameter& 7.36 kpc& $\star$\\
B-band effective radius consistent& \multirow{2}{*}{$r_ {\sfrac{1}{2}}=2.7\pm 0.1$  kpc}& $\multirow{2}{*}{\citealt{hunteretal2001}}$\\
with Lyman-break analogs&&\\
   \hline
  \multicolumn{3}{l}{$\star$: http://annesastronomynews.com/photo-gallery-ii/galaxies-clusters/ngc-2366/}
 \end{tabular}
\end{table*}

\section {Observations and Data Reduction}
\label{Obs}
The observations are performed in three different epochs by using the Russian-Turkish--Telescope (RTT150)\footnote{Specifications of RTT150 and T\"{U}B\.{I}TAK Faint Object Spectrometer and Camera (TFOSC) are available at \href{http://www.tug.tubitak.gov.tr}{http://www.tug.tubitak.gov.tr}.}. The first epoch is on 29th of March 2017 (CCD1), the second epoch is  on 23rd - 24th of December 2017 (CCD2) and the third epoch is  on 9th - 10th September 2018 (CCD2). The seeing conditions were consistent throughout the whole observations  within a range of  1.8$-$2.0 $''$.

\subsection{Imaging observations}
The RTT150 has a Ritchey-Chr\'{e}tien optical system functioning together with Cassegrain and Coude focal systems. We used a TFOSC-CCD, with the f/7.7 focal ratio Cassegrain low resolution faint object spectrograph and camera throughout our observations. Our imaging observations were completed by using H$\alpha$, H$\alpha$ cont., [S\,{\sc ii}] and [S\,{\sc ii}] cont. narrow band filters. The  technical properties of the CCDs (i.e. CCD1 and CCD2) used during our three epochs of observations are given in Table~\ref{Table2}. The log of the optical observations are shown in Table~\ref{Table3}.

\begin{table*}
\centering
 \caption{Specifications of the CCDs on RTT150 Telescope.}
 \label{Table2}
 \begin{tabular}{@{}ccc@{}}
 \hline
Properties & CCD1 & CCD2 \\
\hline
 Pixel Format & 2048 $\times$ 2048 pixel & 2048 $\times$ 2048 pixel\\
 Pixel Sizes & 15 $\mu$m & 13.5 $\mu$m\\
 Image View & 13.3 $\times$ 13.3 arcmin$^{2}$ & 11.3 $\times$ 11.3 arcmin$^{2}$ \\
 \hline
\end{tabular}
\end{table*}

\begin{table*}
\centering
 \caption{The filter characteristics and exposure times.}
 \label{Table3}
 \begin{tabular}{@{}ccccl@{}}
 \hline
Filter & Wavelength & FWHM &Exposure times & Observations date  \\
 & (\AA) & (\AA)& (s)&\\
 \hline
 H$\alpha$ & 6563 & 80 & 3 $\times$ 300 & 29th March 2017 \\
  &  &  & 6 $\times$ 900 & 23rd - 24th December 2017 \\
  &  &  & 6 $\times$ 900  & 9th - 10th September 2018 \\
H$\alpha$ cont. & 6446 & 123 & 1 $\times$ 300 & 29th March 2017\\
  &  &  & 6 $\times$ 300& 23rd - 24th December 2017 \\
  &  &  & 6 $\times$ 300 & 9th - 10th September 2018 \\
$[$S$\,${\sc ii}$]$ & 6728 & 54& 3 $\times$ 300 & 29th March 2017 \\
  &  &  & 6 $\times$ 900  & 23rd - 24th December 2017 \\
  &  &  & 6 $\times$ 900 & 9th - 10th September 2018 \\
$[$S$\,${\sc ii}$]$ cont. & 6964 & 350 & 1 $\times$ 300 & 29th March 2017 \\
  &  &  & 6 $\times$ 300 & 23rd - 24th December 2017 \\
  &  &  & 6 $\times$ 300 & 9th - 10th September 2018 \\
 \hline
\end{tabular}
\end{table*}

\subsection{Imaging data analysis}
The raw data was processed by using an Image Reduction and Analysis Facility ({\sc IRAF})\footnote{developed by USA National Optical Astronomical Observatories (NOAO)- \\https://iraf-community.github.io/} for data reduction  and  also by  The European Southern Observatory Munich Image Data Analysis System {\sc (ESO-MIDAS)}\footnote{http://www.eso.org}.

\begin{figure}
\centering
 \vspace*{17pt}
 \includegraphics[width=11cm]{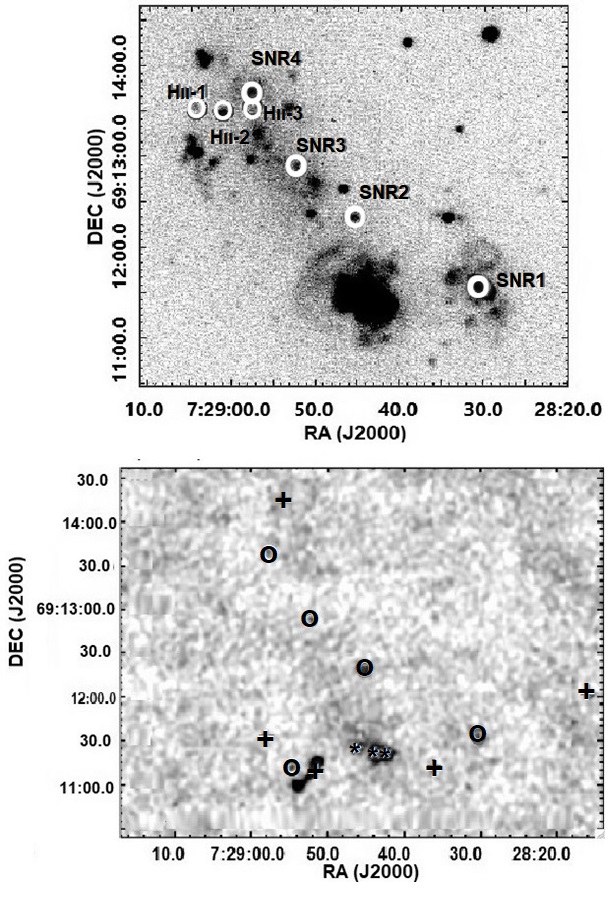}
 \caption{Top Panel: Combined RTT150 data: The H$\alpha$ image of NGC 2366 with J2000. Possible SNR candidates are given with thick white circles with our SNR1, SNR2, SNR3 and SNR4, which are matched with the radio SNR candidates of \citet{chomiukandwilcots2009} are shown. Our H\,{\sc ii} regions are given with thin white circles H\,{\sc ii}-1, H\,{\sc ii}-2 and H\,{\sc ii}-3 are shown. Bottom Panel: \citet{chomiukandwilcots2009} 20 cm radio map of NGC 2366, with J2000, where ``+'' indicates background galaxies, ``circles'' their SNR candidates and ``$\star$'' signs are their H\,{\sc ii} regions.}
 \label{fig:fig1}
\end{figure}

Standard procedures were used for data reduction, which include the bias corrections, overscan at field, and cosmic ray elimination by making use of the {\sc IRAF} CCDPROC package and {\sc MIDAS} CCDRED application packages. To inspect the locations of the stars, positions of red stars from the USNO A2.0 catalog \citet{monetetal1998} were implemented. Fig.~\ref{fig:fig1} presents the positions of the SNR candidates overlaid on the Digitized Sky Survey (DSS)\footnote{https://archive.eso.org/dss/dss} image of the NGC 2366 galaxy. [S\,{\sc ii}]-[S\,{\sc ii}]cont.  and H$\alpha$-H$\alpha$ cont.  images were obtained after achieving the second step from the cleaned images. Finally, the [S\,{\sc ii}]/H$\alpha$ ratio can be reached by forming the ratio of these above mentioned two images. {Standard stars Feige34, HR 5501 and HR 8634 \citep[see e.g.][]{masseyetal1988, hamuyetal1992, hamuyetal1994} were  both observed to calibrate its optical flux.}

We began the selection process of candidate SNRs in NGC 2366 by constructing continuum-subtracted H$\alpha$ and [S\,{\sc ii}] $\lambda \lambda$6716, 6731 images and calculating the [S\,{\sc ii}]/ H$\alpha$ image ratios. The regions having image ratio values $\geq$ 0.4 were identified as candidate SNRs \citep{blairandlong1997}.

Preliminary SNR candidates were found by blinking between continuum-subtracted [S\,{\sc ii}] and H$\alpha$ sub-field images. For visual inspection of the fields to search for candidates, we produced images of a 2$'$ $\times$ 2$'$ region. The bright features in the continuum-subtracted [S\,{\sc ii}] image were checked against the continuum-subtracted H$\alpha$ image to make sure that the stars were not crudely subtracted. If any feature seen on the [S\,{\sc ii}] image looked brighter than it was in the H$\alpha$ image, we tagged it as a candidate SNR, a possible target for follow-up spectral observations.

We made use of the [S\,{\sc ii}], H$\alpha$ images and their continuum-subtracted images to measure the total counts for each candidate SNR. Then a circular aperture in the continuum-subtracted images was chosen to sum the Analogue to Digital Units (ADU) counts. A concentric annulus region was applied to pick the background counts, which were later subtracted from the aperture sum.

The seeing was obtained during interference filter imaging observations (namely, 1.9 $''$ corresponding to $\sim$ 32 pc for the assumed distance to NGC 2366 of 3.44 Mpc), which is implemented in constraining the aperture sizes used for the measurements of fluxes. We did not include radii calculations for the detected SNR candidates, because of the differences we found between seeing and pixel scales. We used data from \citet{cardellietal1989} to correct flux values for interstellar extinction.

\citet{jacobyetal1987} gave a description of the conversion process from instrumental counts into physical units of (erg cm$^{-2}$ s$^{-1}$). Fig.~\ref{fig:fig1} shows the combined three epochs H$\alpha$ image of NGC 2366 with J(2000) coordinates. Detected possible SNR candidates are shown together with their  numbers  of 1, 2, 3, 4. Individual images in each filter were aligned and combined with the \texttt{combine} packages of {\sc IRAF} and {\sc ESO-MIDAS}. The combination of these data sets are appropriate for the particular analysis presented in the paper. During these observations two different CCDs were used as can be seen  from  Table~\ref{Table2}.  CCD1 and CCD2 have  almost identical angular resolution, and their quantum efficiencies are similar to each other as well as their seeing between our three  epochs of observation.

To be able to get  better statistics and therefore better scientific results, we have decided to combine the imaging data for three epochs with the two CCDs. The possible SNR candidates obtained from the combined data are shown in Fig.~\ref{fig:fig1}.

A total of four possible SNR candidates near the central part of the NGC  2366 with appropriate [S\,{\sc ii}]/H$\alpha$ ratio and three  H\,{\sc ii} regions are suggested by us, from the combined data as shown in our Table \ref{tab:Table4} and in  Fig.~\ref{fig:fig1} top panel.

\begin{table*}
\centering
 \caption{Combined optical imaging data for the four SNR candidates (SNR1, SNR2, SNR3, SNR4), their corresponding radio coordinates reported by \citet{chomiukandwilcots2009} of our IDs of SNR1, SNR2, SNR3, SNR4 and three  H\,{\sc ii} regions (1, 2, 3)  with their H$\alpha$ fluxes, [S\,{\sc ii}]/H$\alpha$ ratios and their errors. SNR1, SNR2 and H\,{\sc ii} 1  correspond here to the  two  optical SNR candides and one H\,{\sc ii} region respectively, as reported by  \citet{vuceticetal2019} and are shown as Vuc5, Vuc34 and Vuc 52.}
 \begin{threeparttable}
 \label{tab:Table4}
 \begin{tabular}{@{}lcccc@{}}
\hline\hline
     ID & Optical Coordinates & Radio Coordinates$^{+}$& [S$\,${\sc ii}$]$ / H$\alpha$ & I (H$\alpha$) $\times$ $10^{-15}$\\
     & $RA(J2000)$  & $RA(J2000)$ & & \multirow{2}{*}{(erg cm$^{-2}$ s$^{-1}$)}\\

       & $Dec.(J2000)$ &  $Dec.(J2000)$  & & \\
\hline\hline
 SNR1  & $07^{h} 28^{m} 30^{s}$.70  & $07^{h} 28^{m}  30^{s}$.41 & \multirow{2}{*}{0.68 ($\pm 0.05$)} & \multirow{2}{*}{2.10}\\

(Vuc 5)$^{*}$ &  $\ang[angle-symbol-over-decimal]{69;11;33.60}$ &  $\ang[angle-symbol-over-decimal]{69;11;33.80}$& & \\

(Cho 07)$^{+}$ &  &  & & \\
\hline
\hline

SNR2& $07^{h} 28^{m} 52^{s}$.30  & $07^{h} 28^{m} 52^{s}$.10  & \multirow{2}{*}{0.57 ($\pm 0.04$)} & \multirow{2}{*}{0.36}\\

(Vuc 34)$^{*}$  &  $\ang[angle-symbol-over-decimal]{69;12;53.00}$ & $\ang[angle-symbol-over-decimal]{69;12;54.40}$ & & \\

(Cho 15)$^{+}$  &   &  & & \\
\hline
SNR3 & $07^{h} 28^{m} 57^{s}$.50  &$07^{h} 28^{m} 57^{s}$.67 &  \multirow{2}{*}{0.55 ($\pm 0.02$)} & \multirow{2}{*}{0.14}\\

(Cho 18)$^{\rm +}$&  $\ang[angle-symbol-over-decimal]{69;13;41.80}$ &  $\ang[angle-symbol-over-decimal]{69;13;41.00}$ &  & \\
\hline

SNR4 & $07^{h} 28^{m} 45^{s}$.24  & $07^{h} 28^{m} 45^{s}$.26  & \multirow{2}{*}{ 0.75 ($\pm 0.05$)} & \multirow{2}{*}{0.11}\\

(Cho 12)$^{\rm +}$ &  $\ang[angle-symbol-over-decimal]{69;12;20.70}$ & $\ang[angle-symbol-over-decimal]{69;12;19.80}$ & & \\

\hline\hline

H\,{\sc ii} 1  & $07^{h} 29^{m} 01^{s}$  && \multirow{2}{*}{0.11 ($\pm 0.01$)} & \multirow{2}{*}{0.05}\\

(Vuc$^{\rm *}$52) &  $\ang[angle-symbol-over-decimal]{69;13;30}$ &  & &\\
\hline

\multirow{2}{*}{H\,{\sc ii} 2} & $07^{h} 28^{m} 57^{s}$.50  & &\multirow{2}{*}{0.26 ($\pm 0.01$)} & \multirow{2}{*}{0.13}\\

& $\ang[angle-symbol-over-decimal]{69;13;33}$ & & & \\
\hline

\multirow{2}{*}{H\,{\sc ii} 3} & $07^{h} 29^{m} 04^{s}$.20  & & \multirow{2}{*}{0.09 ($\pm 0.01$)} & \multirow{2}{*}{0.07}\\

&  $\ang[angle-symbol-over-decimal]{69;13;31.05}$  && & \\

\hline\hline
\end{tabular}
\begin{tablenotes}
\item {\bf Notes.} $^{\rm *}$ From \citet {vuceticetal2019},
\item {} $^{\rm +}$ From \citet {chomiukandwilcots2009}.
\end{tablenotes}
\end{threeparttable}
\end{table*}

\subsection{Spectral observations}

Our spectral observations were performed by using the Cassegrain-TFOSC / CCD of RTT150. Spectral measurements were obtained in the $\lambda$3230--9120 {\AA} range with the grism G15. In total spectra of 5 different regions have been analysed. For the flux calibration, we have observed the standard stars (Feige34, HR 5501 and HR 8634) as our spectrophotometric standards as suggested by \citep[see e.g.][]{masseyetal1988, hamuyetal1992, hamuyetal1994}. Fe-Ar calibration lamp frames were obtained for slit width for each observation. High-resolution long-slit spectra of two possible SNR candidates, SNR1 and SNR4, are reported here together with those of three  possible detected H\,{\sc ii} regions (H\,{\sc ii}-1, H\,{\sc ii}-2 and H\,{\sc ii}-3).

\subsection{Spectral data analysis}

The spectroscopic observations were performed on March 29, 2017  (Epoch 1) and  September 9, 2018 (Epoch 3) with the RTT150 using the medium resolution spectrometer TFOSC. The grism G15 with the nominal dispersion of $\sim$ 8 {\AA} pixel$^{-1}$ was used. The reduction and analysis of a spectrum are made using the package Longslit context of {\sc IRAF}. The exposure time for each spectrum  is 3600 s. The slit width we used during our spectral observations was 1.78$''$  (100 $\mu$). The resolution at H$\alpha$ is 16 {\AA}.

The ratio [S\,{\sc ii}]/H$\alpha$ and the electron density are calculated from the [S\,{\sc ii}] ($\lambda$6716/$\lambda$6731) flux ratios following the work done by \citet{osterbrockandferland2006}. Using the [O\,{\sc iii}]$\lambda$5007/H$\beta$ line ratio, the shock velocities, $V_{\rm s}$, proposed for extragalactic SNRs are estimated from the study of \citet{matonickandfesen1997} and the  measurements of [O\,{\sc iii}]$\lambda$5007/H$\beta$ ratios reported by \citet{dopitaetal1984}  providing an estimate for the [O\,{\sc iii}]$\lambda$5007/H$\beta$ ratio as a function of shock velocity. These authors  also provided the plots of this ratio in their Figs 5 and 6 in terms of $V_{\rm s}$. Our [O\,{\sc iii}]$\lambda$5007/H$\beta$ ratio range (2.61$-$3.57) is obtained from our individual spectral analyses for the four SNR candidates and the shock velocities of H$\alpha$ were produced by a dust screen distribution using the relations of \citet{relanoetal2006, buatetal2002} as can be seen from Table \ref{tab:Table5}.

The interstellar extinction and the extinction of H$\alpha$ produced by such a dust screen distribution were also determined through the relations of \citet{relanoetal2006, buatetal2002} together with  Neutral Hydrogen column density  relation from \citet{predehlandschmitt2009}. This is similar to the determinations reported for another nearby galaxy by \citet{ercanetal2018}.

The spectral parameters of SNR1 and SNR4, including the relative line flux and electron density ($N_{\rm e}$)\footnote{calculated with The Space Telescope Science Data Analysis System (STSDAS) task \texttt{nebular.temden} program; this task is based on the program FIVEL \citep{derobetisetal1987, shawanddufour1995} for a five level atom} parameter were derived from the [S\,{\sc ii}]($\lambda$6716/$\lambda$6731) ratios for an assumed electron temperature of 10$^4$ K \citep{osterbrockandferland2006}.

\begin{figure}
\centering
 \vspace*{17pt}
\includegraphics[width=13cm]{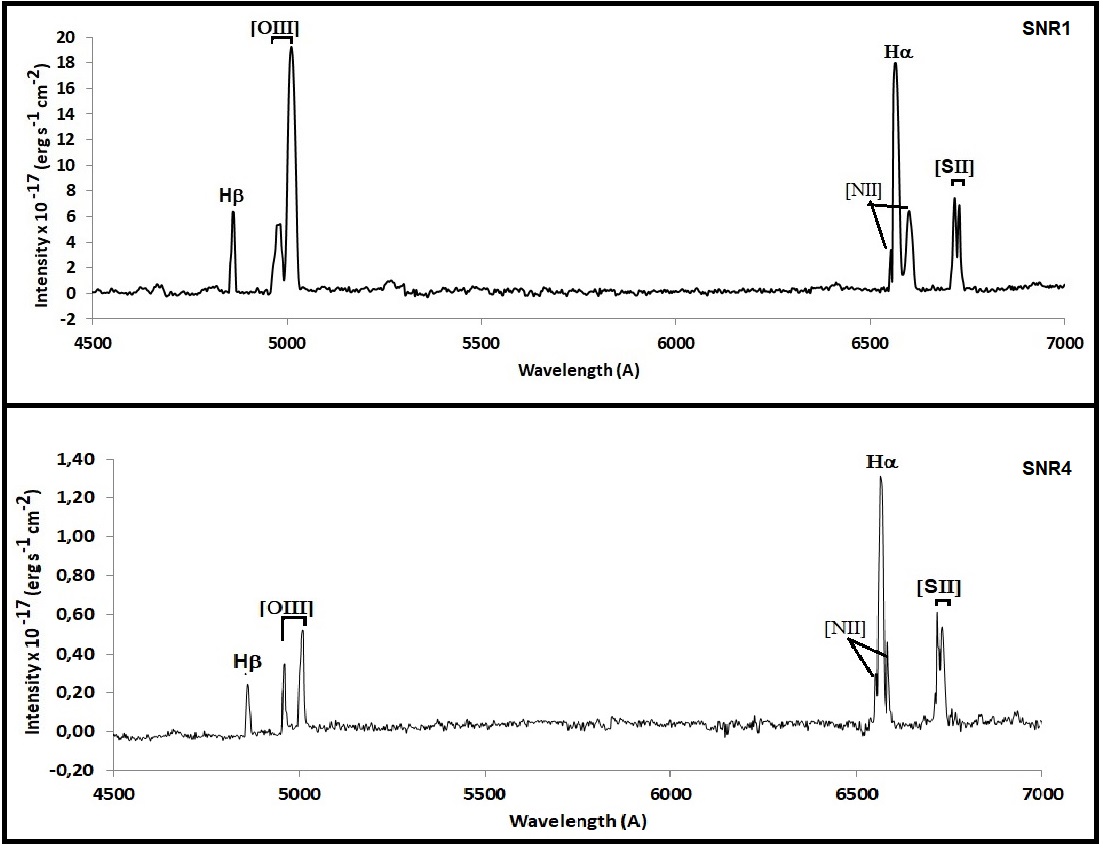}
 \caption{Spectra of SNR1 and SNR4 (Epoch 1) in $\lambda$4500$-$7000 {\AA}  range. Its Balmer H$\alpha$ $\lambda$6563 {\AA}, H$\beta$ $\lambda$4861 {\AA}, forbidden lines [O\,{\sc iii}] $\lambda\lambda$4959,5007 {\AA}, [N\,{\sc ii}] $\lambda\lambda$6548,6584 {\AA}, [S\,{\sc ii}] $\lambda\lambda$6717,6731 {\AA} are shown.}
 \label{fig:fig2}
\end{figure}

\begin{figure}
\centering
 \vspace*{17pt}
\includegraphics[width=13cm]{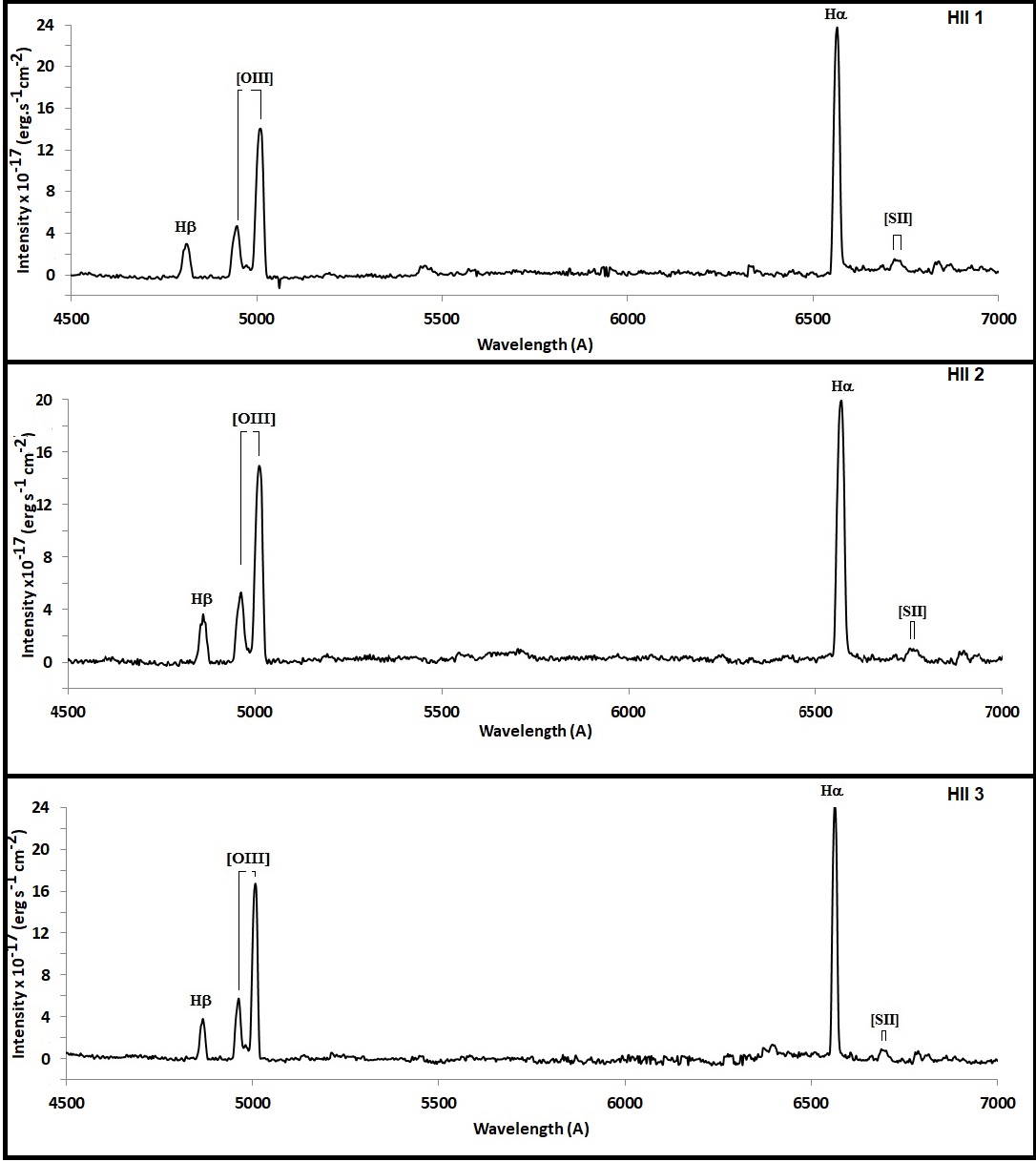}
 \caption{Spectra of H\,{\sc ii} region 1 (Epoch 1), 2 (Epoch 1) and 3 (Epoch 3) candidates in $\lambda$4500$-$7000 {\AA} range.}
 \label{fig:fig3}
\end{figure}

The spectra of our possible candidates SNR1 and SNR4 together with those of the three  possible H\,{\sc ii} regions of 1, 2, and 3 in the range of {$\lambda$4500$-$7000 {\AA}, } are presented in Table \ref{tab:Table5} and Fig.~\ref{fig:fig2} and Fig.~\ref{fig:fig3}. Our results for SNR1 and SNR2 are found to be consistent with the optical observations reported recently  by \citet{vuceticetal2019} . However, our SNR3, SNR4, H\,{\sc ii}-2 and H\,{\sc ii}-3 have not been detected by these authors. The limiting flux sensitivity of our  imaging observations is obtained by  taking a  blanck and faintest part in a circle same circular aperture with our 4 SNRs and found  to be  0.037 $\times$ 10$^{-15}$ erg cm$^{-2}$ s$^{-1}$.   In \citet{drissenetal2000} NGC 2366BG7, located at $RA(J2000)=07^{h} 28^{m} 45^{s}.3$, $Dec.(J2000)=\ang[angle-symbol-over-decimal]{69;12;19.2}$ is reported as one of the background galaxies in the direction of NGC 2366 by using MKR 71 and NGC 2363  Hubble Space Telescope (HST) data. Its position is found to be consistent with our SNR4. It is also  consistent with one of the five radio SNRs suggested by \citet{chomiukandwilcots2009} as their one out of five SNRs so-called Cho 12.

\citet{chomiukandwilcots2009} reported radio observations of NGC 2366 and suggested the sources Cho 07, Cho 15, Cho 18 and Cho 12 to be possible SNR  candidates. In this study, our possible four SNR candidates (see Table \ref{tab:Table4}) are all found to be in a good agreement with their 20 cm radio locations.
Our  H\,{\sc ii} regions reported here are  consistent with the findings of \citet{vuceticetal2019} results.

\begin{table*}
\centering
 \caption{Relative line intensity for the SNR candidates (SNR1 and SNR4). Fluxes are normalised F(H$\alpha$) = 100. The signal-to-noise ratios (S/N) of the emission lines are given together with other physical parameters.}
 \label{tab:Table5}
 \begin{threeparttable}
 \begin{tabular}{@{}lcccc@{}}
 \hline
Lines (\AA) & \multicolumn{4}{c}{Fluxes (H$\alpha$=100) and S/N} \\
\hline
& \multicolumn{2}{c}{SNR1}&\multicolumn{2}{c}{SNR4}\\& F & S/N&  F & S/N
\\\cline{2-5}

H$\beta$ ($\lambda$4861)& 31.36 &5  &18.70 &4\\

$[$O$\,${\sc iii}$]$ ($\lambda$4959) & 36.92& 5  &37.40 &9\\

$[$O$\,${\sc iii}$]$ ($\lambda$5007) & 111.80 & 15 &48.78 &5 \\

$[$N$\,${\sc ii}$]$ ($\lambda$6548) & 28.25&12&24.39 &5\\

H$\alpha (\lambda$ 6563) & 100 &15 &100 &18\\
$[$N$\,${\sc ii}$]$ ($\lambda$6584) & 43.16&6&32.52&6\\

$[$S$\,${\sc ii}$]$ ($\lambda$6716) & 36.97 &5&42.28&4\\

$[$S$\,${\sc ii}$]$ ($\lambda$6731) & 36.56&5&40.65 &4\\
\hline
Parameters $\&$ Line Ratios&\multicolumn{4}{c}{Values}\\
\hline

I (H$\alpha$) ($10^{-15}$ erg cm$^{-2}$ s$^{-1}$) & \multicolumn{2}{c}{1.92 $\pm$ 0.04}&\multicolumn{2}{c}{0.12 $\pm$ 0.03}\\

[S\, {\sc ii}]$^{\rm a}$/ H$\alpha$ & \multicolumn{2}{c}{0.74 $\pm$ {0.11}}&\multicolumn{2}{c}{0.82 $\pm$ 0.08}\\

[S\,{\sc ii}]$\lambda6716/\lambda$6731& \multicolumn{2}{c}{1.01 $\pm$ 0.08}&\multicolumn{2}{c}{1.04 $\pm$ 0.07}\\

[O\,{\sc iii}] ($\lambda$5007)/H$\beta$ ($\lambda$4861) & \multicolumn{2}{c}{3.57 $\pm$ 0.07}&\multicolumn{2}{c}{2.61 $\pm$ 0.14}\\

$V_{\rm s}$ (km s$^{-1}$) &\multicolumn{2}{c}{80--100} &\multicolumn{2}{c}{80--100}\\

$E(B-V)$& \multicolumn{2}{c}{0.36 $\pm$ 0.06}& \multicolumn{2}{c}{0.26 $\pm$ 0.07}\\

$A_{\rm {(H\alpha)}}$ & \multicolumn{2}{c}{0.25 $\pm$ 0.06}&\multicolumn{2}{c}{1.43 $\pm$ 0.07}\\

$N$(H$\,${\sc i}) ($10^{20}$ cm$^{-2}$) & \multicolumn{2}{c}{2.42 $\pm$ 0.05}  & \multicolumn{2}{c}{13.80 $\pm$ 0.2}\\

$N_{\rm e}$(cm$^{-3}$)& \multicolumn{2}{c}{581.91 $\pm$ 168.61} &\multicolumn{2}{c}{513.38 $\pm$ 135.88}\\
 \hline
\end{tabular}
\begin{tablenotes}
\item {\bf Notes.} $^{\rm a}$ [S\,{\sc ii}] is the combination of the $\lambda$6716 and $\lambda$6731 flux values.
\end{tablenotes}
\end{threeparttable}
\end{table*}

 In order to verify the photometric calibration, H$\alpha$ fluxes of our possible SNR candidates are compared and given in our Table \ref{tab:Table5}. Our flux values are found to be in the range of (0.11 - 2.10) $\times$ 10$^{-15}$ erg cm$^{-2}$ s$^{-1}$. \citet{vuceticetal2019}'s work of optical imaging observations revealed the existence of  two optical SNR candidates together with their characteristics of H$\alpha$ and [S\,{\sc ii}] fluxes across the two fields of view in NGC 2366. Our SNR1 and SNR2 locations are consistent with their findings. However, our SNR3 and SNR4  are consistent with the radio results given by \citet{chomiukandwilcots2009} which may indicate the possible existence of two new optical SNR candidates in NGC 2366.

\citet{vuceticetal2019} reported 67 possible H\,{\sc ii} regions in their work and their so-called Vuc 52 is consistent with our H\,{\sc ii}-1 position as shown in our Table \ref{tab:Table4}. The XMM-Newton X-ray position given at $RA(J2000)=07^{h} 28^{m} 58^{s}.2$, $Dec.(J2000)=\ang[angle-symbol-over-decimal]{69;11;34}$ by \citet{thuanetal2014} is found to be very close when compared with 5th radio position reported by \citet{chomiukandwilcots2009}

\section{Discussions and conclusions}
\label{Conclusions}
In this work, we present an optical study for the SNR candidates using narrow-band images and their spectra obtained with RTT150-TUG  for the first time. We have  identified four  possible SNRs, with appropriate [S\,{\sc ii}]/H$\alpha$ ratios and three  possible H\,{\sc ii} regions. Our conclusions are;

\begin{enumerate}{

\item From optical imaging, the ratio [S\,{\sc ii}]/H$\alpha$ $\geq$0.4 is used in our work to identify the possible SNR candidates. We identified a total of four  SNR candidates in NGC 2366 together with a possible existence of three  H\,{\sc ii} regions using our combined  optical data. Out of four SNR candidates reported here, the two of them are newly discovered SNR candidates in NGC 2366. New spectroscopic observations of five of these sources (SNRs \& H\,{\sc ii} regions) are presented as well.

\item Spectral analyses are also performed here for two of our possible SNR candidates in NGC 2366. We report the range of line fluxes and the parameters of [S\,{\sc ii}]/H$\alpha$ = 0.74 - 0.82, [S\,{\sc ii}]($\lambda$6716/$\lambda$6731) = 1.01 - 1.04, shock wave velocities of 80 < $V_{\rm s}$ < 100 km s$^{-1}$, its optical extinction range of $E(B-V)$ = 0.26 - 0.36, $A_{\rm (H\alpha)}$ in the range of 0.25 - 1.43, absorbing column density range of $N$(H\,{\sc i}) = (2.42 - 13.80) $\times10^{20}$ cm$^{-2}$ and electron density range of $N_{\rm e}$ = (513 - 581) cm$^{-3}$, as shown in Table \ref{tab:Table5}. Two out of four SNR candidate locations reported in this work are found to be consistent with  \citet{vuceticetal2019}'s results. We must point out here that  optical flux values  we have reported here are found  to be consistent  with  other nearby galaxy SNRs  (see e.g. \citet{sonbasetal2009, sonbasetal2010}. \citet{ pakmoretal2012, pakmoretal2013, taubenbergeretal2013, leonidakietal2013}  are  mentioned  the spectral features of SNRs in nearby galaxies  and the relation with their emission lines. A viable  model for normally bright  SNe Ia, where nebular O\,{\sc i} emission had never been observed  could be  a similar case that fit with our spectral results presented in this study.

\item According to \citet{chomiukandwilcots2009} radio observations, five radio SNR candidates are reported. All of our possible SNR candidates locations presented here  are found to be  consistent with four of their  SNR positions (see our Table \ref{tab:Table4}).

\item The fainter and possibly extended \textit{XMM-Newton} source J072830.4+691132 reported by \citet{thuanetal2014} coincides with our SNR candidate SNR1 to within about 2 $''$,  which may suggest its association with a possible SNR candidate when it is considered in junction with our spectral results of SNR1.

\citet{stevensandstrickland1998}'s \textit{ROSAT} PSPC study of NGC 2366 supported the existence of characteristic extended region at $RA(J2000)=07^{h} 28^{m} 24^{s}$, $Dec.(J2000)=\ang[angle-symbol-over-decimal]{69;11;00}$ which is not exactly coincident with any of our observed possible SNR candidates, studied here. However, it can be considered  nearest to our  SNR1 candidate at $RA(J2000)=07^{h} 28^{m} 30^{s}$.7, $Dec.(J2000)=\ang[angle-symbol-over-decimal]{69;11;33.60}$.

\item We must emphasize here that in the X-ray band, NGC 2366 has very limited observations and previous X-ray  studies  outlined above could not contribute much to the existence of possible SNR candidates in this nearby galaxy. Further deep observations are needed by satellites such as {\it Chandra} and/or {\it XRISM} to achieve a better understanding of the X-ray properties of NGC 2366. For this purpose we are in a stage of submitting  a {\it Chandra} proposal for a better understanding of the possible SNR candidates and the physics behind them.

\item About the SN rate, \citet{chakrabarti2018} reported the SN rates are obtained using the data from Lick Observatory Supernova Search (LOSS). The rate beyond the optical radius of spiral galaxies host 2.5 $\pm$ 0.5 SNe per millennium. The rates of core-collapse SNe that may collapse to form the massive black holes detected by the Laser Interferometer Gravitational-Wave Observatory (LIGO) in the outer discs of spiral is reported to be 1.5 $\pm$ 0.15 per millennium and for dwarf galaxies is 2.6 $\pm$ 1.5 SNe per millennium or in otherwords, 31,000 $\pm$ 18,000 SNe Gpc$^{-3}$ yr$^{-1}$. The relative ratio of core-collapse to SNe Ia is comparable in the inner and outer parts  in dwarf galaxies. There is no SN rate reported . Since there is no reported SN rate for NGC 2366,  one can assume that  the predicted SNe rate for dwarf galaxies mentioned above may well be applicable  for NGC 2366 }.

\end{enumerate}

\section*{Acknowledgements}
We appreciate the valuable comments given by the  referee and we  thank T\"{U}B\.{I}TAK-TUG for their support in using RTT150 (Russian-Turkish Telescope) for our observations performed through Project number 15BRTT150-842. This work is supported by Bogazici University, BAP with project number 8563 and ENE thanks BAP for their support. We thank Dr. Aytap Sezer for many useful discussions and Prof.Dr.Reynier Peletier for his valuable comments.

\bibliographystyle{plain}

\bibliography{mybibfile}

\onecolumn

\end{document}